\documentclass[11pt]{article}
\pdfoutput=1
\usepackage{amsmath,epsf,amssymb,latexsym,array,color}
\usepackage{graphicx,wasysym}
\usepackage{cancel}

\def\be{\begin{equation}}
\def\ee{\end{equation}}
\def\ba{\begin{array}}
\def\ea{\end{array}}
\newcommand{\beq}{\begin{equation}}
\newcommand{\eeq}[1]{\label{#1}\end{equation}}
\newcommand{\bea}{\begin{eqnarray}}
\newcommand{\eea}[1]{\label{#1}\end{eqnarray}}

\usepackage{ifpdf}
\ifx\pdfoutput\undefined
   \pdffalse
\else
   \pdfoutput=1
   \pdftrue
  \usepackage[pdftex]{hyperref}
\begin{document}


\begin{titlepage}

\hskip 1.5cm

\begin{center}
{\huge \bf{Constraining Liberated Supergravity}}
\vskip 0.8cm  
{\bf \large Hun Jang\footnote{hun.jang@nyu.edu} and Massimo Porrati\footnote{massimo.porrati@nyu.edu}}  
\vskip 0.75cm

{\em Center for Cosmology and Particle Physics\\
	Department of Physics, New York University \\
	726 Broadway, New York, NY 10003, USA}
	
	\vspace{12pt}

\end{center}

\begin{abstract}
Fermionic terms in a class of locally supersymmetric theories called ``liberated supergravity'' 
are nonrenormalizable interactions proportional to 
inverse powers of the supersymmetry breaking scale and  Planck mass $M_{pl}$. This property defines an 
intrinsic cutoff for liberated supergravities, 
which are therefore effective theories valid only below energies that never exceed the cutoff. 
Requiring that the cutoff exceeds current theoretical and observational bounds shows that the new scalar potential terms 
allowed by liberated supergravity  can neither change the cosmological constant predicted by supergravity by 
any observable amount, nor give measurable contributions to particle masses.
We show that it is nevertheless possible to define a simple liberated supergravity model of slow roll inflation valid up to 
energy scales that are well above the Hubble parameter during inflation and exceeds observable limits after inflation. 
The key to
constructing a viable model is to change the supersymmetry breaking scale, from a Planck-scale value during inflation,
to TeV-scale after inflation. 
\end{abstract}

\vskip 1 cm
\vspace{24pt}
\end{titlepage}
\tableofcontents

\section{Introduction}
Supergravity strongly constraints the form of the scalar potential~\cite{cfgvnv} hence it makes the construction of specific 
models of elementary interactions or inflation challenging. The literature on supergravity inflationary potentials 
is vast and somewhat unnecessary for this paper, which is devoted instead to the study of a new model of supergravity, 
called ``liberated 
supergravity.''~\cite{fkr} As its name suggests, the scalar potential of liberated supergravity need not be of the special form
given in~\cite{cfgvnv}; it is in fact a completely general function of the scalar fields. 

As any other gravity theory, liberated supergravity contains nonrenormalizable interaction, so it is by construction an 
effective theory valid only up to a finite cutoff that cannot exceed the (reduced) Planck scale 
$\Lambda_{cut}\lesssim 1/\sqrt{8\pi G} \equiv M_{pl}$.  
Differently from models where supersymmetry is nonlinearly realized this scale can be parametrically larger than the
supersymmetry (SUSY) breaking scale $M_S$. This is also different from the models described 
in~\cite{akk, acik,cftp,ar}, where the cutoff is also only required to be $\mathcal{O}(M_S)$.
 In liberated supergravity, instead, the cutoff can be parametrically larger than $M_S$, because the latter
 is defined through the F and/or D-term expectation values while the former depends on both the new terms 
 present in the ``liberated'' scalar potential $\mathcal{U}$ and the expectation values of the auxiliary fields.
 
For a given value of $\Lambda_{cut}$ the corrections to the scalar potential 
due to the new terms allowed by liberated supergravity are severely constrained by our 
Eq.~\eqref{generic_constraints}, which refine and make quantitative 
Eqs. (3.30)-(3.31) of ref.~\cite{fkr}.  These constraints are the first main result of our paper, because they
show that even in the least restrictive case $\Lambda_{cut}\sim M_S$ the ``liberated'' correction to the vacuum energy 
are completely negligible. The same is true to liberated corrections to particle masses. 

 Another question worth asking is whether liberated supergravity can nevertheless produce a slow-roll inflationary potential 
 with a cutoff well above the Hubble scale (the latter condition is necessary for the self-consistency of an effective 
 theory of inflation). We answer in the affirmative by exhibiting a toy model with a simple, explicit potential that can
 satisfy all 
 constraints. The key feature that makes it possible to describe Hubble scale physics in our model is a 
 change occurring in the supersymmetry breaking potential from one that gives a fixed high value $M_S\gg H$ 
 during slow roll to a no-scale potential after the end of inflation.
 
 Section II succinctly describes the construction of liberated supergravity in the superconformal tensor calculus formalism. 
 Section III describes the construction of its fermionic terms and derives the key inequalities and 
 constraints that any such theory must obey. Section IV presents a toy model of slow-roll inflation in liberates supergravity.

\section{Liberated $\mathcal{N}=1$ supergravity in superconformal tensor calculus}

In this section, we summarize the construction of liberated $\mathcal{N}=1$ supergravity~\cite{fkr} 
using superconformal tensor calculus~\cite{fvp,Linear} that will be discussed in more details in~\cite{jp20}. Liberated supergravity was described in~\cite{fkr} using the superspace formalism, where a K\"{a}hler transformation is introduced to
 remove the variation of the action under a super-Weyl rescaling, while a 
super-Weyl-K\"{a}hler transformation is promoted to an Abelian gauge symmetry to produce the liberated supergravity. In
 the superconformal formalism one instead introduces a conformal compensator multiplet, called  $S_0$, 
 which eliminates the variation
  while keeping the K\"{a}hler potential invariant under superconformal symmetry. Hence, differently from
   the superspace case,  we need to define such a gauge transformation independently of superconformal symmetry.

To do this, it is useful to write the invariant actions in the two different formalisms~\cite{kyy}:
\begin{eqnarray}
&& [\mathcal{V}]_D = 2 \int d^4\theta E \mathcal{V}, ~ [\mathcal{S}]_F = \int d^2\theta \mathcal{E}\mathcal{S} + \int d^2 \bar{\theta} \bar{\mathcal{E}} \bar{\mathcal{S}},
\end{eqnarray}
where $\mathcal{V}$ is a composite 
superconformal real multiplet with the Weyl/chiral weights (2,0), $\mathcal{S}$ is a composite superconformal chiral 
multiplet with Weyl/chiral weights (3,3) while $E$ and $\mathcal{E}$ are the corresponding D/F-term measure 
 densities~\cite{fvp}. 
The action must be invariant under a super-Weyl-K\"ahler transformation that shifts
 the superspace densities as $ E \rightarrow Ee^{2\Sigma + 2\bar{\Sigma}}$ and $\mathcal{E} \rightarrow \mathcal{E} e^{6\Sigma}$. This requires that the corresponding superconformal multiplets transform as 
\be
\mathcal{V} \rightarrow \mathcal{V} e^{-2\Sigma-2\bar{\Sigma}},\qquad  \mathcal{S} \rightarrow \mathcal{S} e^{-6\Sigma}.
\ee
To describe the liberated supergravity in the superconformal formalism, we therefore promote the super-Weyl-K\"{a}hler transformation to a gauge symmetry under which the compensator is inert, so that the transformation rules are given by
\begin{eqnarray}
&& K \rightarrow K + 6\Sigma+6\bar{\Sigma},\quad W \rightarrow We^{-6\Sigma},\quad \bar{W} \rightarrow \bar{W} e^{-6\bar{\Sigma}}, \nonumber\\
&& T \rightarrow e^{-4\Sigma+2\bar{\Sigma}}T,\quad \bar{T} \rightarrow e^{2\Sigma-4\bar{\Sigma}}\bar{T}, \quad S_0 \rightarrow S_0\nonumber\\
&& \mathcal{W}_{\alpha} \rightarrow e^{-3\Sigma}\mathcal{W}_{\alpha},\quad T(\bar{\mathcal{W}}^2) \rightarrow T(\bar{\mathcal{W}}^2) e^{-4\Sigma -4\bar{\Sigma}},
\end{eqnarray}
where $\Sigma$ is a chiral superfield, $W$ is the superpotential, $\mathcal{W}_{\alpha}$ is the field strength
 of a real vector multiplet and $T$ is the chiral projection.
With these rules, a Lagrangian of the liberated $\mathcal{N}=1$ supergravity equivalent to that of Ref. \cite{fkr} is:
\begin{eqnarray}
 \mathcal{L}_{NEW}  \equiv \bigg[ \mathcal{Y}^{-2}\frac{ \mathcal{W}^2(K)\bar{\mathcal{W}}^2(K)}{T(\bar{w}^2(K))\bar{T}(w^2(K)) }\mathcal{U}(Z^I,\bar{Z}^{\bar{\imath}})\bigg]_D.
\label{NewLagrangian}
\end{eqnarray}

In Eq~\eqref{NewLagrangian}, we have introduced the notations\footnote{In superconformally invariant theories there are no dimensionless parameters and the Planck scale is introduced by the super-Weyl gauge choice $\Upsilon \equiv s_0s_0^*e^{-K/3}=M_{pl}^2$, where $\Upsilon$ is by definition the lowest component of $\mathcal{Y}$.} $\mathcal{Y} \equiv (S_0\bar{S}_0e^{-K(Z^I,\bar{Z}^{\bar{\imath}})/3})$, $w^2(K) \equiv \mathcal{W}^2(K)/\mathcal{Y}^2$, $\bar{w}^2(K) \equiv \bar{\mathcal{W}}^2(K)/\mathcal{Y}^2$; we denoted with $\mathcal{U}(Z^I,\bar{Z}^{\bar{\imath}})$ a generic 
function of the matter chiral multiplets $Z^I$'s and with $K(Z^I,\bar{Z}^{\bar{\imath}})$ 
the supergravity K\"{a}hler potential. We also call $\mathcal{W}_{\alpha}(K)$ the field strength multiplet corresponding to 
the K\"{a}hler potential. 
By denoting with $(w,c)$ the Weyl/chiral weights of a multiplet we have the following assignment: 
$(1,1)$ for $S_0$, $(2,0)$ for $\mathcal{Y}$, $(3/2,3/2)$ for $\mathcal{W}_{\alpha}(K)$, $(-1,3)$ for $w^2(K)$, and
$(0,0)$ for $Z,K(Z^I,\bar{Z}^{\bar{\imath}}),T(\bar{w}^2(K)),\bar{T}(w^2(K)),\mathcal{U}(Z^I,\bar{Z}^{\bar{\imath}})$. By assuming that 
$\mathcal{U}(z^I,\bar{z}^{\bar{\imath}})$ is inert under the gauge symmetry and using 
$w^2  \rightarrow  w^2 e^{-2\Sigma+4\bar{\Sigma}}$, $\bar{T} (w^2) \rightarrow \bar{T}(w^2)$, we see that the whole
Lagrangian is invariant under the gauge symmetry, so that it reproduces liberated supergravity.

Next we find the bosonic contribution to the scalar potential. We define a composite 
superconformal multiplet $\mathbb{N}$
made of the superconformal chiral multiplets $Z^i \equiv (z^i,P_L\chi^i,F^i)$ ($i = 0,I,W,T$), which can be $(z^0\equiv s_0,P_L\chi^0,F^0)$, $(z^I,P_L,\chi^I)$, $(z^W \equiv W,P_L\chi^W,F^W)$\footnote{Here, $W(K)$ denotes the lowest component of the field strength multiplet $\mathcal{W}^2(K)$. It is {\em not} the usual superpotential 
$W$.}, $(z^T \equiv T(\bar{w}^2),P_L\chi^T,F^T)$. The lowest component $N$ of $\mathbb{N}$, is given by
\begin{eqnarray}
N &\equiv& (s_0s_0^*e^{-K(z^I,\bar{z}^{\bar{\imath}})/3})^{-2}\frac{ W\bar{W}}{T(\bar{w}^2)\bar{T}(w^2) }\mathcal{U}(z^I,\bar{z}^{\bar{\imath}}),\nonumber\\{} \label{def_of_newterm}
\end{eqnarray}
we get the component Lagrangian by using the D-term formula. It reads as follows: 
\begin{eqnarray}
 && \mathcal{L}_{NEW} \equiv [\mathbb{N}]_De^{-1} \nonumber\\
 &&= N_{i\bar{\jmath}} \bigg( -\mathcal{D}_{\mu}z^i\mathcal{D}^{\mu}\bar{z}^{\bar{\jmath}} - \frac{1}{2} \bar{\chi}^i \cancel{\mathcal{D}} \chi^{\bar{\jmath}} - \frac{1}{2} \bar{\chi}^{\bar{\jmath}}\cancel{\mathcal{D}}\chi^i + F^i\bar{F}^{\bar{\jmath}}\bigg)
\nonumber\\
&& +\frac{1}{2}\bigg[ N_{ij\bar{k}} \Big( -\bar{\chi}^i\chi^j \bar{F}^{\bar{k}} + \bar{\chi}^i (\cancel{\mathcal{D}}z^j)\chi^{\bar{k}} \Big) +h.c. \bigg]
\nonumber\\&&+ \frac{1}{4}N_{ij\bar{k}\bar{l}} \bar{\chi}^i\chi^j \bar{\chi}^{\bar{k}}\chi^{\bar{l}}
\nonumber\\
&& +\bigg[\frac{1}{2\sqrt{2}}\bar{\psi}\cdot \gamma 
\bigg( N_{i\bar{\jmath}} F^i \chi^{\bar{\jmath}} - N_{i\bar{\jmath}} \cancel{\mathcal{D}}\bar{z}^{\bar{\jmath}}\chi^i -\frac{1}{2}N_{ij\bar{k}} \chi^{\bar{k}}\bar{\chi}^i\chi^j\bigg)  \nonumber\\
&& +\frac{1}{8} i\varepsilon^{\mu\nu\rho\sigma} \bar{\psi}_{\mu} \gamma_{\nu} \psi_{\rho} \bigg( N_i \mathcal{D}_{\sigma}z^i + \frac{1}{2} N_{i\bar{\jmath}} \bar{\chi}^i \gamma_{\sigma} \chi^{\bar{\jmath}} \nonumber\\
&&+ \frac{1}{\sqrt{2}} N_i \bar{\psi}_{\sigma} \chi^i  \bigg)+h.c. \bigg] 
+\frac{1}{6} N \left( -R(\omega) +\frac{1}{2} \bar{\psi}_{\mu} \gamma^{\mu \nu\rho} R'_{\nu\rho}(Q) \right)
\nonumber\\
&&-\frac{1}{6\sqrt{2}} \left( N_i\bar{\chi}^i + N_{\bar{\imath}}\bar{\chi}^{\bar{\imath}} \right)\gamma^{\mu\nu} R'_{\mu\nu}(Q),\label{componentLagrangian}
\end{eqnarray}
where $N_{i\bar{\jmath}}$, $N_{ij\bar{k}}$, and $N_{ij\bar{k}\bar{l}}$ are the derivatives with respect to $z^i,z^{\bar{\imath}}$ for $i,j = 0,I,W,T$: $N_i\equiv \partial N /\partial z^i$ etc..  The gravitino is denoted by $\psi$ and other details will be given 
in~\cite{fvp}. As for the detailed structure of the fermions $\chi^i\equiv P_L\chi^i$, we find~\cite{jp20}
\begin{eqnarray}
P_L\chi^i&=&
\begin{cases}
P_L\chi^0,\\
P_L\chi^I,\\
P_L\chi^W = 4 \tilde{\mathcal{F}} K_{\bar{\imath}J}[(\cancel{\partial}z^{J})P_R\chi^{\bar{\imath}}-\bar{F}^{\bar{\imath}}P_L\chi^J ] \\ 
\qquad\qquad+ \cdots  +7~\textrm{fermion terms},\\
P_L \chi^T = \frac{1}{\Upsilon^2}\bigg[~  4 (\cancel{\partial}\tilde{\mathcal{F}}) K_{\bar{\imath}J}[(\cancel{\partial}z^{J})P_R\chi^{\bar{\imath}}-\bar{F}^{\bar{\imath}}P_L\chi^J]
 \\\qquad\qquad- \Big(\frac{2}{s_0^*}\cancel{\partial}s_0^* -\frac{2}{3}K_{\bar{K}}\cancel{\partial}\bar{z}^{\bar{K}} - 2\gamma^{\mu}(b_{\mu}+iA_{\mu})\Big)
 \\ \qquad\qquad\quad  \times 4 \tilde{\mathcal{F}} K_{\bar{\imath}J}[(\cancel{\partial}z^{J})P_R\chi^{\bar{\imath}}-\bar{F}^{\bar{\imath}}P_L\chi^J] ~ \bigg]\\
 \qquad\qquad+ \cdots + 9~\textrm{fermion terms}.
\end{cases}\nonumber\\{}
\end{eqnarray}
where $\tilde{\mathcal{F}} =  2K_{I\bar{\jmath}} \left( -\partial_{\mu} z^I \partial^{\mu} \bar{z}^{\bar{\jmath}} +F^I\bar{F}^{\bar{\jmath}}\right)$ and 
\beq
 \cancel{\partial} \tilde{\mathcal{F}}= 2(\cancel{\partial}K_{I\bar{\jmath}}) \left( -\partial_{\mu} z^I \partial^{\mu} \bar{z}^{\bar{\jmath}} +F^I\bar{F}^{\bar{\jmath}}\right)
+2K_{I\bar{\jmath}} \left( -(\cancel{\partial}\partial_{\mu} z^I) \partial^{\mu} \bar{z}^{\bar{\jmath}}-\partial_{\mu} z^I (\cancel{\partial}\partial^{\mu} \bar{z}^{\bar{\jmath}}) +(\cancel{\partial}F^I)\bar{F}^{\bar{\jmath}}+F^I(\cancel{\partial}\bar{F}^{\bar{\jmath}})\right). 
\eeq{8}

Notice that $\chi^i$ contains not only the fundamental fermions $\chi^0$ and $\chi^I$ but also two composite 
chiral fermions $\chi^W$ and $\chi^T$.

A straighforward use\footnote{$\mathcal{L}_B \supset N_{i\bar{\jmath}}F^i\bar{F}^{\bar{\jmath}} \sim N_{W\bar{W}}F^W\bar{F}^{\bar{W}} = \Upsilon^{-2}\frac{1}{\mathcal{C}_{T}\mathcal{C}_{\bar{T}}}\mathcal{U} F^W\bar{F}^{\bar{W}}
= \Upsilon^{-2}\frac{\Upsilon^2\Upsilon^2}{\bar{F}^{\bar{W}}F^W}\mathcal{U} F^W\bar{F}^{\bar{W}} = \Upsilon^2 \mathcal{U} \equiv V_{NEW}$ where we have used $\mathcal{C}_{T} =- \frac{1}{2}\mathcal{K}_{\bar{w}} \sim -\frac{1}{2} \frac{\mathcal{K}_{\bar{W}}}{\Upsilon^2} =-\frac{1}{2} \frac{(-2\bar{F}^{\bar{W}})}{\Upsilon^2} = \frac{\bar{F}^{\bar{W}}}{\Upsilon^2}$ and $\mathcal{C}_{\bar{T}} = \frac{F^W}{\Upsilon^2}$. 
and $\mathcal{C}_T$ is the lowest component of the superconformal multiplet of $T$.}  of the superconformal tensor calculus then gives the scalar potential in the form 
\begin{eqnarray}
V_{NEW} =  (s_0s_0^*e^{-K/3})^{2}  \mathcal{U}(z^I,\bar{z}^{\bar{\imath}}).
\end{eqnarray}

The super-Weyl gauge choice $\Upsilon = s_0s_0^*e^{-K/3}= M_{pl}^2 \equiv 1$ puts the action in the Einstein frame and reproduces 
the scalar potential of ref.~\cite{fkr}
\begin{eqnarray}
V_{NEW} = \mathcal{U}(z^I,\bar{z}^{\bar{\imath}}).
\end{eqnarray}
This then implies that the total scalar potential generically is
\begin{eqnarray}
V = V_D + V_F + V_{NEW},
\end{eqnarray}
where $V_D,V_F$ are the usual D/F-term potentials. The additional contribution to the scalar potential, $V_{NEW}$, is 
an arbitrary function of the $z^I$s; since it does not obey any constraint it fully justifies the name ``liberated'' for 
the new class of supergravities introduced in~\cite{fkr}.

\section{Fermionic terms in liberated $\mathcal{N}=1$ supergravity}

In this section, we investigate the fermionic terms in  liberated $\mathcal{N}=1$ supergravity in the superconformal formalism\footnote{The detailed derivation will be done in~\cite{jp20}.}. First of all, focusing only on matter couplings, i.e. looking at terms independent of $\psi$, we read the following terms from Eq.~\eqref{componentLagrangian}
\begin{eqnarray}
\mathcal{L}_{F1} &\equiv&  -N_{i\bar{\jmath}}\mathcal{D}_{\mu}z^i\mathcal{D}^{\mu}\bar{z}^{\bar{\jmath}}\Big|_{\psi=0},\quad 
\mathcal{L}_{F2} \equiv  -\frac{1}{2} N_{i\bar{\jmath}} \bar{\chi}^i \cancel{\mathcal{D}} \chi^{\bar{\jmath}}\Big|_{\psi=0}  ,\nonumber\\
\mathcal{L}_{F3} &\equiv&   -N_{i\bar{\jmath}} F^i\bar{F}^{\bar{\jmath}} \Big|_{\psi=0},\quad 
\mathcal{L}_{F4} \equiv -\frac{1}{2} N_{ij\bar{k}} \bar{\chi}^i\chi^j \bar{F}^{\bar{k}}\Big|_{\psi=0},\nonumber\\
\mathcal{L}_{F5} &\equiv& \frac{1}{2}  N_{ij\bar{k}} \bar{\chi}^i (\cancel{\mathcal{D}}z^j)\chi^{\bar{k}}\Big|_{\psi=0} ,\nonumber\\
 \mathcal{L}_{F6} &\equiv&  \frac{1}{4}N_{ij\bar{k}\bar{l}} \bar{\chi}^i\chi^j \bar{\chi}^{\bar{k}}\chi^{\bar{l}} \Big|_{\psi=0} ,\quad 
\mathcal{L}_{F7} \equiv - \frac{N}{6}R(\omega)|_{\psi=0}.\label{fermi_Lagrangian}
\end{eqnarray}
Here, we observe that the fermionic terms in the  effective Lagrangian contain couplings to the function $\mathcal{U}$ 
and its derivatives since $N \propto \mathcal{U}$.  

The general structure of the fermionic terms can be found as a power series in derivatives of the composite multiplet $N$ (i.e. $N_i,N_{i\bar{\jmath}},N_{ij\bar{k}}$ and $N_{ij\bar{k}\bar{l}}$). The $r$-th derivative of $N$, denoted with 
$N_{i\ldots\bar{l}}^{(r)}$ has the following generic form
\begin{eqnarray}
N_{i\ldots\bar{l}}^{(r)} &=& N^{(r=q+p+m+k)}_{q,p,m,k} 
= (\partial_0^q\partial_I^{(k-n)}\Upsilon^{-2})\Upsilon^{4+2m}\frac{\mathcal{U}^{(n)}}{\tilde{\mathcal{F}}^{4+2m}}W^{1-p_1}\bar{W}^{1-p_2}
\nonumber\\
&=&\Big(
(-1)^{q_1+q_2} (q_1+1)!(q_2+1)! s_0^{-(2+q_1)} {s_0^*}^{-(2+q_2)}(\partial_I^{(k-n)}e^{2K/3})\Big) \Upsilon^{4+2m}\frac{\mathcal{U}^{(n)}}{\tilde{\mathcal{F}}^{4+2m}}W^{1-p_1}\bar{W}^{1-p_2}.
\nonumber\\{} \label{Gen_str_N_deriv}
\end{eqnarray}
where 
\begin{eqnarray}
 W &=&  -2K_{\bar{\imath}J}[\bar{\chi}^{J}(\overline{\cancel{\mathcal{D}}\bar{z}^{\bar{\imath}}})-\bar{F}^{\bar{\imath}}\bar{\chi}^{J}]K_{{\bar{\imath}}'J'}[(\cancel{\mathcal{D}}\bar{z}^{{\bar{\imath}}'})\chi^{J'}-F^{J'}\chi^{{\bar{\imath}}'}]  - K_{{\bar{\imath}}J}[\bar{\chi}^{J}(\overline{\cancel{\mathcal{D}}\bar{z}^{\bar{\imath}}})-\bar{F}^{\bar{\imath}}\bar{\chi}^{J}] K_{\bar{\imath}'\bar{\jmath}'K'}[\chi^{K'}\bar{\chi}^{\bar{\imath}'}\chi^{\bar{\jmath}'}] \nonumber\\
&& - K_{\bar{\imath}\bar{\jmath}K}[\bar{\chi}^{\bar{\jmath}}\chi^{\bar{\imath}}\bar{\chi}^{K}]K_{{\bar{\imath}}'J'}[(\cancel{\mathcal{D}}\bar{z}^{{\bar{\imath}}'})\chi^{J'}-F^{J'}\chi^{{\bar{\imath}}'}]  - \frac{1}{2}K_{\bar{\imath}\bar{\jmath}K}[\bar{\chi}^{\bar{\jmath}}\chi^{\bar{\imath}}\bar{\chi}^{K}]K_{\bar{\imath}'\bar{\jmath}'K'}[\chi^{K'}\bar{\chi}^{\bar{\imath}'}\chi^{\bar{\jmath}'}],\\
\bar{W} &=& (W)^*,
\end{eqnarray}
$\tilde{\mathcal{F}} \equiv  2K_{I\bar{\jmath}} \left( -\partial_{\mu} z^I \partial^{\mu} \bar{z}^{\bar{\jmath}} +F^I\bar{F}^{\bar{\jmath}}\right)$; 
$\mathcal{U}^{(n)}$ ($0 \leq n \leq 4$) is the $n$-th derivative of the 
function $\mathcal{U}(z^I,\bar{z}^{\bar{\imath}})$ with respect to $z^I,\bar{z}^I$, 
which are the lowest component of the matter chiral  multiplets; $q=q_1+q_2$ where $q_1$ 
($q_2$) is the order of the derivative w.r.t. the compensator scalar $s_0$ ($s_0^*$); $p=p_1+p_2$ where $p_1$ ($p_2$) is the order of the derivative w.r.t. the field strength multiplet scalar $W$ ($\bar{W}$); $m=m_1+m_2$ where $m_1$ ($m_2$) is the 
order of the derivative w.r.t. the chiral projection multiplet scalar $T(\bar{w}^2)$ ($\bar{T}(w^2)$); $k$ is the order of the derivative w.r.t. the matter multiplet scalar $z^I$; 
$n$ is the order of the derivative acting on the new term $\mathcal{U}$ w.r.t. the matter multiplet; $q$ is the total order of 
derivative w.r.t. the compensator scalars $s_0$ and $s_0^*$. 

To find restrictions on $V_{NEW}$ coming from fermionic terms, we have to identify the most singular terms in the 
Lagrangian. These terms can be found using the fact that powers of $\tilde{\mathcal{F}}$ in the denominator may lead to a 
singularity which gets stronger when $m$ increases by taking more derivatives with respect to the lowest component of the
 multiplet $T(\bar{w}^2)$ as seen from Eq.~\eqref{Gen_str_N_deriv}. Hence, we will investigate the fermionic terms containing only derivatives with respect to the chiral projection
 and matter scalar indices, i.e. $T$ and $I$, in order to find the terms coupled to $\mathcal{U}^{(n)}$ that contain the 
 maximal inverse powers of $\tilde{\mathcal{F}}$. They are those with $q=p=0$ and $k=n$. We note in particular that 
 if our theory has a single chiral matter multiplet then the most singular terms are 
 found to be the couplings to the derivatives proportional to $N_{T\bar{T}}$, $N_{WT\bar{T}}$, $N_{W\bar{W}T\bar{T}}$
 while for two or more chiral matter multiplets they are $N_{TT\bar{T}\bar{T}}$. The latter terms vanish identically for a 
 single multiplet because of Fermi statistics.
 
First of all, let us examine the single matter chiral multiplet case. Due to Fermi statistics, the possible fermionic terms are 
proportional only to three tems, $\mathcal{U}^{(0)}$, $\mathcal{U}^{(1)}$, and $\mathcal{U}^{(2)}$, so that the maximal
 order of the derivative with respect to the chiral projection that can appear in the Lagrangians scalar is two and
 appears in the terms 
 proportional to $N_{T\bar{T}}$, $N_{WT\bar{T}}$, and $N_{W\bar{W}T\bar{T}}$. To show that such terms do not 
 vanish consider 
\begin{eqnarray}
\mathcal{L}_{F2}|_{q=p=0,k=n} &\supset&   \Upsilon^{2+2m}\frac{\mathcal{U}^{(n)}}{\tilde{\mathcal{F}}^{4+2m}}W\bar{W}  \bigg[-\frac{1}{2} \Big((\bar{\chi}^I)^{n_1} (\Upsilon^{-2} 4 (\cancel{\partial}\tilde{\mathcal{F}}) K_{\bar{\imath}J}(\cancel{\partial}z^{J})\bar{\chi}^{\bar{\imath}}P_R)^{m_1}\Big)
\nonumber\\
&&\times
\Big((\cancel{\mathcal{D}}\chi^I)^{n_2} (\cancel{\mathcal{D}}(\Upsilon^{-2} 4 (\cancel{\partial}\tilde{\mathcal{F}}) K_{\bar{\imath}J}(\cancel{\partial}z^{J})P_R\chi^{\bar{\imath}}))^{m_2}\Big)\bigg]_{\psi=0},
\end{eqnarray}
where $m = m_1 + m_2$, $n = n_1 + n_2$, and $2=m+n$. Restoring the mass dimensions by fixing the super-Weyl gauge\footnote{Here, we use the convention of the superconformal formalism that all physical bosonic and fermionic matter
 fields have dimensions 0 and 1/2 respectively and  $\mathcal{F}$ has dimension 2 while the compensator 
 $s_0$ has dimension 1~\cite{fvp}. Through dimensional analysis, we find $[\mathcal{D}_{\mu}]=1,[z^i]\equiv 0+[i],[\chi^i]\equiv \frac{1}{2}+[i],[F^i]\equiv 1+[i]$ where $i=0,I,W,T$ and
 $[0]=1,[I]=0,[W]=3,[T]=0$.} (i.e. $\Upsilon = M_{pl}^2$, $s_0=s_0^*=M_{pl} e^{K/6}$, $P_L\chi^0 = \frac{1}{3}s_0 K_I P_L\chi^I= \frac{1}{3}M_{pl} e^{K/6} K_I P_L\chi^I$, and $b_{\mu}=0$), we obtain
\begin{eqnarray}
\mathcal{L}_{F2}|_{q=p=0,k=n} &\supset&   M_{pl}^{2(2+2m)}\frac{\mathcal{U}^{(n)}}{\tilde{\mathcal{F}}^{4+2m}} W\bar{W}  \bigg[-\frac{1}{2} \Big((\bar{\chi}^I)^{n_1} (M_{pl}^{-4} 4 (\cancel{\partial}\tilde{\mathcal{F}}) K_{\bar{\imath}J}(\cancel{\partial}z^{J})\bar{\chi}^{\bar{\imath}}P_R)^{m_1}\Big)
\nonumber\\
&&\times
\Big((\cancel{\mathcal{D}}\chi^I)^{n_2} (M_{pl}^{-4}( 4\cancel{\mathcal{D}} (\cancel{\partial}\tilde{\mathcal{F}}) K_{\bar{\imath}J}(\cancel{\partial}z^{J})P_R\chi^{\bar{\imath}}))^{m_2}\Big)\bigg]_{\psi=0}
\approx M_{pl}^{4} \frac{\mathcal{U}^{(n)}}{\tilde{\mathcal{F}}^{4+2m}} \mathcal{O}_F^{(\delta)},
\end{eqnarray}
where we require $m_a+n_a = 1$ for $a=1,2$ since we are studying the second derivative term $N_{i\bar{\jmath}}$ coupled to $\bar{\chi}^i$ and $\cancel{\mathcal{D}}\chi^{\bar{\jmath}}$. 
Redefining $\tilde{\mathcal{F}}$ to be dimensionless by $\tilde{\mathcal{F}} \rightarrow M_{pl}^2 \tilde{\mathcal{F}}$, we obtain
\begin{eqnarray}
 \mathcal{L}_{F2}|_{q=p=0,k=n} \supset M_{pl}^{-4-2m} \frac{\mathcal{U}^{(n)}}{\tilde{\mathcal{F}}^{4+2m}} 
 \mathcal{O}_F'^{(\delta)}. 
\end{eqnarray}
We then find $\delta = 8+4m$ by trivial dimensional analysis because the Lagrangian has mass dimension 4.
Then, since $2 = m+ n$, we find that the most singular term is  
\begin{eqnarray}
 \mathcal{L}_{F2}|_{q=p=0,k=n} \supset M_{pl}^{2(n-4)} \frac{\mathcal{U}^{(n)}}{\tilde{\mathcal{F}}^{2(4-n)}} \mathcal{O}_F'^{(2(6-n))}.\label{SingleChiralMatterCase} 
\end{eqnarray}

Next we consider the general case with several multiplets. We shall focus on the fourth derivative term denoted by $N_{ij\bar{k}\bar{l}}$, which gives a four-fermion term. Also, we have to consider the four-fermion product made only of 
the 
chiral fermions with $i=0,I,T$ because they do not contribute one 
power of the F-term $\tilde{\mathcal{F}}$ in the numerator, which would reduce the number of  inverse power of the F-term
 $\tilde{\mathcal{F}}$. This is because the overall factor of $\chi^W$ contains such linear dependence. 
 The effective fermionic Lagrangian \eqref{fermi_Lagrangian} reads then as follows:
\begin{eqnarray}
 \mathcal{L}_{F6}|_{q=p=0,k=n}
&\supset& \Upsilon^{2+2m}\frac{\mathcal{U}^{(n)}}{\tilde{\mathcal{F}}^{4+2m}} W\bar{W}
 \Big(\frac{1}{4}(\chi^I)^n(4 \Upsilon^{-2} (\cancel{\partial}\tilde{\mathcal{F}}) K_{\bar{\imath}J}(\cancel{\partial}z^{J})P_R\chi^{\bar{\imath}})^m\Big).
\end{eqnarray}
After the super-Weyl gauge fixing  we obtain 
\begin{eqnarray}
 \mathcal{L}_{F6}|_{q=p=0,k=n}
&\supset& M_{pl}^{2(2+2m)}\frac{\mathcal{U}^{(n)}}{\tilde{\mathcal{F}}^{4+2m}} W\bar{W}
 \Big(\frac{1}{4}(\chi^I)^n(4 M_{pl}^{-4} (\cancel{\partial}\tilde{\mathcal{F}}) K_{\bar{\imath}J}(\cancel{\partial}z^{J})P_R\chi^{\bar{\imath}})^m\Big) \nonumber\\
&\approx& c M_{pl}^{4}\frac{\mathcal{U}^{(n)}}{\tilde{\mathcal{F}}^{4+2m}} \mathcal{O}_F^{(\delta)}. \nonumber\\{}
\end{eqnarray}
where $ \mathcal{O}(1)\lesssim c \lesssim \mathcal{O}(10^3)$. After doing the same dimensional analysis as in the 
single-multiplet case, we obtain $\delta = 8+4m$ and
\begin{eqnarray}
\mathcal{L}_{F6}|_{q=p=0,k=n} \supset c M_{pl}^{-4-2m}\frac{\mathcal{U}^{(n)}}{\tilde{\mathcal{F}}^{4+2m}} \mathcal{O}_F'^{(8+2m)}.
\end{eqnarray}
Then, since $4 = m+ n$, we can write the most singular terms as 
\begin{eqnarray}
\mathcal{L}_{F6}|_{q=p=0,k=n} \supset cM_{pl}^{2(n-6)}\frac{\mathcal{U}^{(n)}}{\tilde{\mathcal{F}}^{2(6-n)}} \mathcal{O}_F'^{(2(8-n))},\label{GeneralChiralMatterCases}
\end{eqnarray}
 
 Since we are going to use liberated supergravity to describe time-dependent backgrounds such as slow-roll inflation
  we need to look more closely at the structure of $\tilde{\mathcal{F}}$. From its definition 
 $\tilde{\mathcal{F}} \equiv  2K_{I\bar{\jmath}} \left( -\partial_{\mu} z^I \partial^{\mu} \bar{z}^{\bar{\jmath}} +F^I\bar{F}^{\bar{\jmath}}\right)$, 
 we find $\tilde{\mathcal{F}} \equiv  2K_{I\bar{\jmath}} \left( \dot{z}^I\dot{\bar{z}}^{\bar{\jmath}} +F^I\bar{F}^{\bar{\jmath}}\right) >0$ 
 whenever spatial gradients can be neglected. We see that the most singular behaviors of the fermionic terms arises 
 when $\dot{z}^I=0$. By expanding $\tilde{F}$ around a static vacuum and 
 reserving the notation $\mathcal{F}$ for the expectation value $\left<K_{I\bar{\jmath}}F^I\bar{F}^{\bar{\jmath}}\right>$, the 
 effective Lagrangian can finally be rewritten as
 \begin{itemize}
    \item For the  single chiral matter multiplet case, 
    \begin{eqnarray}
     \mathcal{L}_{F2}|_{q=p=0,k=n} \supset  M_{pl}^{2(n-4)} \frac{\mathcal{U}^{(n)}}{\mathcal{F}^{2(4-n)}} \mathcal{O}_F'^{(2(6-n))}.\label{SingleChiralMatterCase_Vac}
    \end{eqnarray}
    \item For two or more chiral matter multiplets, 
    \begin{eqnarray}
     \mathcal{L}_{F6}|_{q=p=0,k=n} \supset c'M_{pl}^{2(n-6)}\frac{\mathcal{U}^{(n)}}{\mathcal{F}^{2(6-n)}} \mathcal{O}_F'^{(2(8-n))}.\label{GeneralChiralMatterCases_Vac}
    \end{eqnarray}
    where $ \mathcal{O}(10^{-2}) \lesssim c' \lesssim  \mathcal{O}(1)$.
\end{itemize}

The effective operators we obtained are generically nonzero even after considering possible cancellations due to Fermi 
statistics or nonlinear field redefinitions. As an example we can take terms containing $\chi^i$. They are made of two 
composite
 chiral multiplets $\chi^W$ and $\chi^T$ and these produce terms that do not vanish on shell (i.e. imposing $\cancel{\partial}P_L\chi^I\approx 0$ for matter fermions). 
For instance, in a theory with only one matter chiral multiplet $(z,P_L\chi,F)$, we have 
$W = -2K_{\bar{z}z}[\{(\cancel{\partial}z)^2- F^*\cancel{\partial}z\}(\bar{\chi}P_R\chi)+\{{F^*}^2-F^*\cancel{\partial}z\} 
(\bar{\chi}P_L\chi) ]+2K_{\bar{z}z}K_{\bar{z}\bar{z}z}(\cancel{\partial}z-F^*)(\bar{\chi}P_L\chi) (\bar{\chi}P_R\chi)$, and 
$\bar{W} = (W)^*$, so that $W\bar{W} = 4K_{\bar{z}z}^4(|\cancel{\partial}z|^2+|F|^2)||\cancel{\partial}z-F^*|^2(\bar{\chi}
P_L\chi)(\bar{\chi}P_R\chi)$. Hence, looking at the possible fermionic terms from $\mathcal{L}_{F1}$, when $i=z,\bar{\jmath}
=\bar{z}$ (i.e. $q=p=m=0,k=n=2$), we get 
\begin{eqnarray}
\mathcal{L}_{F1} &\supset&  \Upsilon^2\frac{\mathcal{U}^{(2)}}{\tilde{\mathcal{F}}^4}W\bar{W}(\partial_{\mu} z\partial^{\mu}\bar{z})
= \Upsilon^2\frac{\mathcal{U}^{(2)}}{\tilde{\mathcal{F}}^4}4K_{\bar{z}z}^4(|\cancel{\partial}z|^2+|F|^2)||\cancel{\partial}z-F^*|^2(\partial_{\mu} z\partial^{\mu}\bar{z})(\bar{\chi}P_L\chi)(\bar{\chi}P_R\chi) . \nonumber\\
\end{eqnarray}
 It is easy to see that this operator does not vanish on the mass shell of the matter scalars, $\square z \approx 0$. 
As another example, from $\mathcal{L}_{F2}$ we get terms containing up to three matter fermions when we consider $q=m=p_1=0,k=n=1,p_2=p=1$
\begin{eqnarray}
\mathcal{L}_{F2} &\supset& \Upsilon^2\frac{\mathcal{U}^{(1)}}{\tilde{\mathcal{F}}^4} \frac{1}{2} W \bar{\chi} \cancel{\partial}P_R\chi^{\bar{W}} \approx  \Upsilon^2\frac{\mathcal{U}^{(1)}}{\tilde{\mathcal{F}}^4} 4K_{\bar{z}z}^2(\cancel{\partial}\tilde{\mathcal{F}})(\cancel{\partial}z)(\bar{\chi}P_R\chi) P_L(\cancel{\partial}z-F^*)^2\chi|_{\mbox{\small 3-fermion terms}}
+\cdots.\nonumber\\{}
\end{eqnarray}

Back to the results in Eqs. \eqref{SingleChiralMatterCase_Vac} and \eqref{GeneralChiralMatterCases_Vac}, the general effective Lagrangians can be cast in the form 
\begin{eqnarray}
\mathcal{L}_F = \Lambda_{cut}^{4-\delta}\mathcal{O}_F'^{(\delta)}= 
\begin{cases}
M_{pl}^{2(n-4)} \dfrac{\mathcal{U}^{(n)}}{\mathcal{F}^{2(4-n)}} \mathcal{O}_F'^{(\delta  = 2(6-n))}~\qquad \textrm{for}~N_{mat} =1,\\
 c'M_{pl}^{2(n-6)}\dfrac{\mathcal{U}^{(n)}}{\mathcal{F}^{2(6-n)}} \mathcal{O}_F'^{(\delta = 2(8-n))} ~\qquad\textrm{for}~N_{mat} \geq 2.
\end{cases}
\end{eqnarray}
where $ \mathcal{O}(10^{-2}) \lesssim c' \lesssim \mathcal{O}(1)$; $N_{mat}$ and $\Lambda_{cut}$ are defined to be the number of chiral multiplets of matter, and the cutoff scale of our effective theory, respectively.

If we demand that our effective theory describe physics up to the energy scale  $\Lambda_{cut}$ we obtain the following
 inequalities:
\begin{eqnarray}
 \mathcal{U}^{(n)} \lesssim \begin{cases}
  \mathcal{F}^{2(4-n)} \left(\dfrac{M_{pl}}{\Lambda_{cut}}\right)^{2(4-n)}  \quad \textrm{where}\quad0\leq n \leq 2\quad\textrm{for}~N_{mat} =1,\\
\mathcal{F}^{2(6-n)}\left(\dfrac{M_{pl}}{\Lambda_{cut}}\right)^{2(6-n)}  \quad \textrm{where}\quad0\leq n \leq 4\quad\textrm{for}~N_{mat} \geq 2.
  \end{cases}\label{pre_inequality}
\end{eqnarray}

A conventional definition of the supersymmetry breaking scale $M_S$ is in terms of F-term expectation value so we define 
$M_S^4=M_{pl}^4\mathcal{F}$, so the constraints on $\mathcal{U}^{(n)}$ become 
\begin{eqnarray}
 \mathcal{U}^{(n)} \lesssim \begin{cases}
  \left(\dfrac{M_S}{M_{pl}}\right)^{8(4-n)}\left(\dfrac{M_{pl}}{\Lambda_{cut}}\right)^{2(4-n)} \quad \textrm{where}\quad0\leq n \leq 2\quad\textrm{for}~N_{mat} =1,\\
\left(\dfrac{M_S}{M_{pl}}\right)^{8(6-n)}\left(\dfrac{M_{pl}}{\Lambda_{cut}}\right)^{2(6-n)} \quad\textrm{where}\quad0\leq n \leq 4\quad\textrm{for}~N_{mat} \geq 2.
  \end{cases}\label{generic_constraints}
\end{eqnarray}

Equation~\eqref{generic_constraints} is the crucial one in our paper, as it constrains precisely the new 
function $\mathcal{U}$ 
introduced by liberated $\mathcal{N}=1$ supergravity. The constraint depends on the reduced Planck scale $M_{pl}$, the
supersymmetry breaking scale $M_S$, $\Lambda_{cut}$ and the number of chiral multiplets of matter in the theory. 
Of course, when we push both the cutoff and supersymmetry breaking scales to the reduced Planck scale, i.e. $\Lambda_{cut} \sim M_S \sim M_{pl}$, we obtain a model-independent universal constraint
\begin{eqnarray}
 \forall n:~ \mathcal{U}^{(n)} \lesssim 1. 
\label{Universal_Constraint1}
\end{eqnarray}

A model where supersymmetry is broken at the Planck scale is hardly the most interesting. In the more interesting case that 
$M_S\ll M_{pl}$ we need the constraints \eqref{generic_constraints} again, so we need to first determine how many matter 
chiral multiplets we have in our theory. The constraints will then depend only on our choice  of 
$\Lambda_{cut}$ and $M_S$. 

In the rest of this section we will examine the constraints in two cases. The first is the true, post-inflationary vacuum of the
 theory. To make a supergravity theory meaningful we want it to be valid at least 
 up to energies $\Lambda_{cut}\gtrsim M_S$. 
The second is slow-roll inflation. In this case we must have $\Lambda_{cut}\gtrsim H$, with $H$ the Hubble constant 
during inflation.

For the post-inflationary vacuum the interesting regime is when $M_S$ is relatively small, say 
$M_S\sim 10\,\mathrm{TeV} \approx 10^{-15} M_{pl}$ and the effective theory is valid up to an energy scale not smaller 
than $M_S$, i.e. $\Lambda_{cut}\gtrsim M_S$. If $\Lambda_{cut} < M_S$ liberated supergravity would be a useless 
complication, since in its domain of validity supersymmetry would be always nonlinearly realized. 
In the post-inflationary vacuum, for the single matter chiral multiplet case, the constraints~\eqref{generic_constraints} thus give for
\bea
&& \mathcal{U}^{(0)} \lesssim \left(\frac{M_S}{M_{pl}}\right)^{32}\left( \frac{M_{pl}}{M_S} \right)^{8} \implies \mathcal{U}^{(0)} \sim 10^{-360},\\
&& \mathcal{U}^{(1)} \lesssim \left(\frac{M_S}{M_{pl}}\right)^{24}\left( \frac{M_{pl}}{M_S} \right)^{6}\implies \mathcal{U}^{(1)} \sim 10^{-270},\\
&& \mathcal{U}^{(2)} \lesssim \left(\frac{M_S}{M_{pl}}\right)^{16}\left( \frac{M_{pl}}{M_S} \right)^{4}\implies \mathcal{U}^{(2)} \sim 10^{-180}.
\eea{Option1_Const}
Notice that $\mathcal{U}^{(3)},\mathcal{U}^{(4)}$ are not restricted. For two or more matter chiral multiplets, the
constraints  are given by
\bea
&& \mathcal{U}^{(0)} \lesssim \left(\frac{M_S}{M_{pl}}\right)^{48}\left( \frac{M_{pl}}{M_S} \right)^{12} \implies \mathcal{U}^{(0)} \sim 10^{-540},\\
&& \mathcal{U}^{(1)} \lesssim \left(\frac{M_S}{M_{pl}}\right)^{40}\left( \frac{M_{pl}}{M_S} \right)^{10}\implies \mathcal{U}^{(1)} \sim 10^{-450},\\
&& \mathcal{U}^{(2)} \lesssim \left(\frac{M_S}{M_{pl}}\right)^{32}\left( \frac{M_{pl}}{M_S} \right)^{8}\implies \mathcal{U}^{(2)} \sim 10^{-360}.
\eea{Option2_Const}
From the constraints on 
$\mathcal{U}^{(0)}$ and $\mathcal{U}^{(2)}$, we find that the liberated scalar potential contributes only a negligibly small 
cosmological constant and negligibly small corrections to the mass terms 
of the chiral multiplet scalars. For the single chiral multiplet case, restoring dimensions we get a vacuum energy density 
\beq
\mathcal{U}^{(0)} \lesssim 10^{-360} M_{pl}^4
\eeq{vac-en}
and scalar masses
\beq
M_z \lesssim M_{pl} \sqrt{|\mathcal{U}^{(2)}|} = 10^{-90}M_{pl}.
\eeq{masses}
These constraints become even tighter if the theory contains more than one chiral multiplet, but the ones we obtained are
already so stringent as to rule out any observable contribution to the cosmological constant and scalar masses from the
new terms made possible by liberated supergravity. We can say that  Eqs.~(\ref{vac-en},\ref{masses}) already send 
liberated supergravity back to prison after the end of inflation.

The constraints during inflation instead can be easily satisfied if during inflation the supersymmetry breaking scale is 
very high, say $M_S = M_{pl}$. In that case, $\mathcal{U}^{(0)} \lesssim \mathcal{O}(1)$. 
After inflation the ``worst case scenario'' constraints coming from Eq.~\eqref{generic_constraints} with $N_{mat}\geq 2$ and
$M_S = 10^{-15}M_{pl}$ are 
\begin{eqnarray}
 \forall n:~\mathcal{U}^{(n)} \lesssim 10^{-120(6-n)}.\label{Option2_Const_after_inf}
\end{eqnarray}

A simple way to satisfy all these constraints is to choose a no scale structure for the supersymmetric part of the scalar potential. This ensures the vanishing of the F-term contribution to the potential independently of the magnitude of the 
F-terms.
\begin{eqnarray}
V_F =e^G(G_IG^{I\bar{\jmath}}G_{\bar{\jmath}}-3) =0,\label{No_scale_structure}
\end{eqnarray}
The total scalar potential is then given by $V = V_D+V_{NEW}$. Thanks to the no-scale structure, we can have 
both $M_S\sim M_{pl}$ and $\mathcal{U}^{(0)} \sim H^2 \sim 10^{-10}$ during inflation. 

Our scenario has $M_S = M_{pl}$ during inflation and $M_S = 10^{-15}M_{pl}$ at the true vacuum in the post-inflation phase, so we see that to satisfy all constraints a transition between the two different epochs must occur, for which the scale of the composite F-term $\mathcal{F}$ changes from $\mathcal{O}(M_{pl})$ to $\mathcal{O}(10^{-15}M_{pl})$.

\section{A minimal model of single-field and slow-roll inflation in liberated $\mathcal{N}=1$ supergravity}

In the previous sections we have argued that liberated $\mathcal{N}=1$ supergravity can be an effective field theory for 
describing the inflationary dynamics while at the same time satisfying all the constraints if a transition that changes the 
supersymmetry breaking scale at the end of inflation is allowed. 
Note that due to the no-scale structure, the scalar potential is given only by an eventual D-term supersymmetric potential 
$V_D$ and the ``liberated'' term $\mathcal{U}$. In this Section we present an explicit minimal model of single-field, 
slow-roll inflation in 
 liberated $\mathcal{N}=1$ supergravity which obeys the inequality $H \ll \Lambda_{cut} = M_{pl} =1$. 

To begin with, let us consider a chiral multiplet $T$ with K\"{a}hler potential  $K(T,\bar{T}) = -3\ln[T+\bar{T}]$ 
and a constant superpotential $W_0$. Then, the supergravity G-function~\cite{cfgvnv} is given by 
\begin{eqnarray}
G \equiv K + \ln |W|^2 = -3\ln[T+\bar{T}] + \ln W_0 + \ln \bar{W}_0.\label{sugra}
\end{eqnarray}
It automatically produces a no-scale structure in which the F-term potential vanishes identically: $V_F =0$. 

Next, let us find the canonically normalized degrees of freedom of the theory. From the kinetic term corresponding to the G
 function~\eqref{sugra}, we read
\bea
\mathcal{L}_K &=& \frac{3}{(T+\bar{T})^2}\partial T\partial \bar{T} \nonumber\\
&=& \frac{3}{4(\textrm{Re}T)^2}(\partial \textrm{Re}T)^2 + \frac{3}{4(\textrm{Re}T)^2}(\partial \textrm{Im}T)^2 \nonumber\\
&=& \frac{1}{2} (\partial \chi)^2 + \frac{1}{2} e^{-2\sqrt{2/3}\chi}(\partial \phi)^2,
\eea{can-nor}
where we have used the following field redefinition 
\begin{eqnarray}
T = \textrm{Re}T + i \textrm{Im}T = \frac{1}{2}e^{\sqrt{2/3}\chi} + i\frac{\phi}{\sqrt{6}}.\label{field_redefinition}
\end{eqnarray}
Note that the $\mathbb{Z}_2$ symmetry $\chi \rightarrow -\chi$ is already explicitly broken by the kinetic Lagrangian, while 
the symmetry $\phi\rightarrow -\phi$ is unbroken. However, even the latter symmetry will be broken by the inflationary 
potential. The field $\chi$ is always canonically normalized while $\phi$ has a canonical kinetic term only at  $\chi =0$.

The composite 
F-term is given by $\mathcal{F} = e^GG_TG^{T\bar{T}}G_{\bar{T}}$ after solving the equation of motion for the auxiliary 
fields $F^I$. For our G function we obtain an exponentially decreasing function 
$\mathcal{F} = 3|W_0|^2/(T+\bar{T})^3= 3|W_0|^2e^{-3\sqrt{2/3}\chi}$. This is what we want to get a viable 
supersymmetry breaking mechanism. The reason
is that we look for a supersymmetry breaking scale during inflation  $M_S^i\sim M_{pl}=1$, while the final scale should be 
parametrically lower than the Planck scale --for instance $M_S^f = 10^{-15}M_{pl}$. To achieve this large difference of
scales, the vacuum expectation value of the field $\chi$ should change during the phase transition. 
On the other hand, the cutoff scale of our model can remain $\mathcal{O}(M_{pl})$ both before and after the 
phase transition. 

We will achieve this with a potential that changes from ($\phi \neq 0, \chi=0$) during inflation to 
$\chi \neq 0, \phi =0$ after inflation. We will also choose $\phi$ as the inflaton field and $\chi$ as the field that controls the 
supersymmetry breaking scale. 

A function $\mathcal{U}$ producing a correct inflationary dynamics is
\beq
\mathcal{U} \equiv \alpha (1-e^{-\sqrt{2/3}\phi})^2(1+\frac{1}{2}\sigma\chi^2),
\eeq{infl-pot}
where $\alpha,\beta,\gamma,\sigma$ are arbitrary positive constants. 

Next, we assume that the mass $m_{\chi}$ of $\chi$ is greater than the Hubble scale $H$ during inflation; 
this is necessary to 
describe a single-field slow-roll inflation governed only by the inflaton field $\phi$. Hence, we impose that during inflation
$m_{\chi}^2 = \alpha \sigma \gg H^2$. Since $\alpha \sim H^2 \sim 10^{-10}$, the condition reduces to 
$\sigma \gg 1$. 

We must also analyze the vacuum structure of the potential. First of all, we explore the minima with respect to $\chi$. 
By computing $\frac{\partial \mathcal{U}}{\partial \chi} = 0$ and defining
 $ V_{\textrm{inf}}  \equiv \alpha (1-e^{-\sqrt{2/3}\phi})^2$ we find that during inflation 
 (where $\phi \neq 0$) the equation of motion for $\chi$ is given by 
 $\sigma \chi V_{\textrm{inf}}  = 0$ 
 so it gives a unique minimum at $\chi =0$. On the other hand after inflation we have $\phi = 0$ and  $V_{\textrm{inf}}=0$, 
 so the equation of motion gives a flat potential in $\chi$. The final position of the field $\chi$ is then determined either
 by the initial conditions on $\chi$ or by small corrections to the either the liberated supergravity potential $\mathcal{U}$ 
 or to $V_F$. We will describe the explicit forms of such corrections in a forthcoming paper~\cite{jp20}. Here we content
 ourselves with pointing out that the simple potential~\eqref{infl-pot} already achieves the goal of making the final 
 supersymmetry breaking scale different from $M_S^i$. 
   
Before studying supersymmetry breaking we notice that a deformation of the scalar potential such as 
$\mathcal{U}$ was obtained using an off-shell \underline{linear} realization of supersymmetry in~\cite{Linear}.  
Therefore, for the new term 
to be consistent, supersymmetry must be broken as usual by some nonvanishing auxiliary field belonging to
the standard chiral 
multiplets and moreover the K\"{a}hler metric of the scalar manifold must be positive~\cite{Linear}. 
So, in spite of the presence of the
new term $\mathcal{U}$, the analysis of supersymmetry breaking is completely standard.  
Since the supersymmetry breaking scale $M_S$ comes from the positive potential part $V_+$, as shown in the 
Goldstino SUSY transformation $\delta_{\epsilon} P_Lv = \frac{1}{2}V_+ P_L \epsilon$ that is constructed with the fermion 
shifts with respect to the auxiliary scalar contributions \cite{fvp}, we have
\begin{eqnarray}
V_+ = e^{G}G_TG^{T\bar{T}}G_{\bar{T}}=\frac{3|W_0|^2}{(T+\bar{T})^3}  = 3|W_0|^2e^{-3\sqrt{2/3}\chi}.\nonumber\\{}
\end{eqnarray}

During inflation we demand that the initial supersymmetry breaking scale is $M_{pl}$, so we identify $W_0 \equiv \dfrac{(M_S^i)^2}{\sqrt{3}}$ and therefore $V_+ = (M_S^i)^4 e^{-3\sqrt{2/3}\chi}$. 
Because $\chi=0$ during
inflation we indeed have $V_+|_{\chi=0,\phi\neq0} = (M_S^i)^4=1  \gg H^2 = \mathcal{O}(10^{-10}M_{pl}^2)$. 

On the other hand, we want to get a much smaller SUSY breaking scale $M_S^f \approx 10^{-15}M_{pl}$ 
around the true vacuum at the 
end of inflation. Thus, at the true vacuum (i.e. $\chi = C$ and $\phi=0$) where $\mathcal{U} = 0$, we get $ V_+|_{\chi=C,\phi=0} \approx (M_S^i)^4 e^{-3\sqrt{2/3}C} \equiv (M_S^f)^4 $. From this, we find where the location of the true vacuum in the $\chi$ direction should be (recall that $\chi$ is a flat direction after inflation) 
\begin{eqnarray}
 C = \sqrt{\frac{8}{3}} \ln \frac{M_S^i}{M_S^f},
\end{eqnarray}
where $M_S^f$ is a free parameter, which we set to be approximately $10^{-15}$ in Planck units. 

The proposed potential $\mathcal{U}$ vanishes after inflation hence it already trivially satisfies the constraints (\ref{generic_constraints}).  So all we need to do is to check that it also satisfies~\eqref{Universal_Constraint1}.
Using $\mathcal{F} = e^{-3\sqrt{2/3}\chi}$, which gives $M_S^i = M_{pl}=1$ during inflation ($\chi =0$), we first have $\mathcal{U}^{(n)}|_{\chi=0} \ll e^{-3m\sqrt{2/3}\chi}\mathcal{O}(1)|_{\chi=0} = \mathcal{O}(1)$. Using Eq. (\ref{field_redefinition}), we find $\partial_T = \sqrt{6}(-i\partial_{\phi} + e^{-\sqrt{2/3}\chi}\partial_{\chi})$ and 
$\partial_{\bar{T}} = \sqrt{6}(i\partial_{\phi} + e^{-\sqrt{2/3}\chi}\partial_{\chi})$. Note that 
$\mathcal{U}^{(n)}|_{\chi=0} \equiv \partial_T^k \partial_{\bar{T}}^l \mathcal{U}(T,\bar{T})|_{\chi=0}$ 
where $n = k+l$. In particular, since the functional dependence
 on $\chi$ does not produce any singularity at $\chi =0$, it is sufficient to check that 
 $\partial_{\phi}^n \mathcal{U} \ll \mathcal{O}(1)$. Thus, because the dependence on $\phi$ is solely given by the 
 Starobinsky inflationary potential, i.e. $V \sim \alpha (1-e^{-\sqrt{2/3}\phi})$, we will get that its derivatives are always less 
 than the coefficient $\alpha$, thanks to the decreasing exponential factor $e^{-\sqrt{2/3}\phi}$.  This implies that the 
 constraint is automatically satisfied since $\alpha \sim 10^{-10} < \mathcal{O}(1)$. So, all consistency conditions can be 
 satisfied by a liberated supergravity potential.


 \subsection*{Acknowledgments} 
 We would like to thank Alexios Kehagias for useful discussions.
 M.P.\ is supported in part by NSF grant PHY-1915219.

\end{document}